\documentclass[aps,pra,reprint,onecolumn,notitlepage,eqsecnum,floatfix,showkeys,nofootinbib]{revtex4-2}

\usepackage{amsmath}
\usepackage{amsbsy}
\usepackage{xcolor}
\usepackage{graphicx}
\usepackage{hyperref}
\usepackage{multirow}
\usepackage{linearb}
\usepackage{relsize}
\usepackage{calrsfs}
\DeclareMathAlphabet\mathbfcal{OMS}{cmsy}{b}{n}

\newcommand{\bq}{\begin{eqnarray}}
\newcommand{\eq}{\end{eqnarray}}
\newcommand{\bqn}{\begin{eqnarray*}}
\newcommand{\eqn}{\end{eqnarray*}}
\newcommand{\bqs}{\begin{subequations}}
\newcommand{\eqs}{\end{subequations}}
\newcommand{\bw}{\begin{widetext}}
\newcommand{\ew}{\end{widetext}}
\newcommand{\xx}{{\boldsymbol x}}

\newcommand{\kk}{{\boldsymbol k}}
\newcommand{\nn}{{\boldsymbol n}}
\newcommand{\rr}{{\boldsymbol r}}

\newcommand{\pp}{{\boldsymbol p}}

\newcommand{\cald}{{\cal D}}

\newcommand{\calp}{{\cal P}}
\newcommand{\cali}{{\cal I}}
\newcommand{\calh}{{\cal H}}

\newcommand{\calm}{{\cal M}}

\newcommand{\red}[1]{{#1}}
\newcommand{\green}[1]{{#1}}

\begin{document}
\title{Edwards Localization}

\author{Riccardo Fantoni}
\email{riccardo.fantoni@scuola.istruzione.it}
\affiliation{Universit\`a di Trieste, Dipartimento di Fisica, strada
  Costiera 11, 34151 Grignano (Trieste), Italy}

\date{\today}

\begin{abstract}
We study the localization problem in quantum stochastic mechanics.
We start from the Edwards model for a particle in a bath of scattering 
centers and prove static localization of the ground state wavefunction
of the particle in a one dimensional square well coupled to Dirac delta 
like scattering centers in arbitrary but fixed positions. We see how
the localization increases for increasing coupling $g$ \green{and increasing
number of scattering centers at constant density}. Then we choose the
scattering centers positions as pseudo random numbers with a uniform 
probability distribution and observe an increase in the localization 
of the average of the ground state over the many positions realizations. 
We discuss how this averaging procedure is consistent with a picture 
of a particle in a Bose-Einstein condensate of of non interacting boson 
scattering centers interacting with the particle with Dirac delta 
functions pair potential. We then study the dynamics of the ground 
state wave function. We conclude with a discussion of the affine 
quantization version of the Lax model which reduces to a system of 
contiguous square wells with walls in arbitrary positions independently
of the coupling constant $g$.
\end{abstract}


\maketitle
\section{Introduction}
\label{sec:intro}

In a recent work \cite{Fantoni25z} we compared the localization property
of a short range polaron \cite{Frohlich1950,Fantoni12d,Fantoni13a} and of 
the (Wick rotated) Anderson problems \cite{Anderson1958}. In
particular we showed how it is possible to study the thermal properties
of the electron in a ionic crystal background, for the polaron problem, or 
of a particle with stochastic kinetic energy interacting with a background 
of scattering sites, for the Anderson problem, through a path integral 
description, after integrating out the degrees of freedom of the background. 

Here we study a third kind of localization that occurs in the Edwards 
model of a particle with deterministic kinetic energy interacting with 
a background of scattering centers with stochastic positions after 
integrating out the degrees of freedom of the background. As 
the above two examples the washing out of the degrees of freedom 
of the background can be treated with Monte Carlo Path Integral methods
in real or imaginary time. Here we will stick to real time. 

While in the polaron problem both the particle and the background have a 
deterministic description, in the Anderson model the (kinetic energy of the) 
particle has a stochastic description, and in the Edwards model the 
(positions of the scattering centers forming the) background has a 
stochastic description. Nonetheless in all three cases we observe 
a localization phase transition of the particle interacting with the
background. In particular here we will see how, in the Edwards model, 
the localization occurs on a circle independently of the strength $g$ 
of the coupling between particle and background and of the switching on 
or off of the disorder. Increasing $g$ or switching on aleatoricity of the 
background scattering centers positions simply increases (make more sharp) 
the localization of the particle ground state, static or dynamic, 
wavefunction.

In order to justify the Monte Carlo integration of the wavefunction 
of the particle and the background respect to the background degrees 
of freedom (the scattering centers positions) we will consider a 
background made of non interacting bosons scattering centers in their 
Bose-Einstein condensed phase \cite{Rose2022}.

\section{Edwards model}
\label{sec:Edwards}

Consider \cite{Rose2022} a $d$-dimensional system made of a particle of 
mass $m$ in a periodic box of volume $\Omega=L^d$ interacting with $N$ 
free spinless bosonic scattering centers of mass $m_c$ at a number 
density $\rho=N/\Omega$ and temperature $T<T_C$ with 
\bq
T_C=\frac{2\pi\hbar^2}{m_ck_B}\left[\frac{\rho}{\zeta(d/2)}\right]^{2/d},
\eq
the critical temperature for Bose-Einstein condensation, where $\hbar$
is Planck constant, $k_B$ is Boltzmann constant, and $\zeta$ is Riemann 
zeta function. Moreover let $v(\rr)$ be the pairwise interaction potential
between the particle and the centers according to {\sl Edwards model}
\cite{Edwards1958}.

Then the wave function of the whole system will be 
$\Psi(\xx;\rr_1,\rr_2,\ldots,\rr_N)$ where $\xx$ is the position of the
particle and $\{\rr_i\}=(\rr_1,\rr_2,\ldots,\rr_N)$ are the positions
of the $N$ scattering centers. If we neglect the interaction between
the $N$ bosons and the particle we may write the normalized wave function 
of the centers as follows
\bq
\Phi(\rr_1,\rr_2,\ldots,\rr_N)=\frac{1}{\sqrt{N!}}
{\rm perm}||\phi_j(\rr_i)||,
\eq 
in terms of the permanent of the $N$ normalized wave functions of each 
center 
\bq
\phi_j(\rr_i)=\frac{1}{\sqrt{\Omega}}e^{i\kk_j\cdot\rr_i},
\eq
with $\kk_j=2\pi\nn_j/L$ and $\hbar\omega_j=(\hbar k_j)^2/2m_c$ his
energy. Here $\nn$ is a $d$-dimensional vector with integer components.
Now below the critical temperature $T_C$ the $N$ centers will undergo
condensation into the $\nn_j={\boldsymbol 0}$ state. So that we will
have
\red{
\bq \label{eq:phibec}
\Phi(\rr_1,\rr_2,\ldots,\rr_N)\propto\sum_\calp\int\cdots\int_\Omega
\prod_{j=1}^N\delta^{(d)}(\rr_{\calp j}-\rr_j')\,
d\rr_1' d\rr_2'\cdots d\rr_N',
\eq
}
where $\delta^{(d)}$ is the Dirac $d$-dimensional delta function and
$\calp$ is a permutation of the $N$ particle indexes.
In general we may expand the wave function $\Psi$ into a basis of 
product states $\psi(\xx)\Phi(\{\rr_i\})$. We may also define
\bq \nonumber
\widetilde{\Psi}(\xx)&=&\langle\Psi(\xx;\rr_1,\rr_2,\ldots,\rr_N)\rangle\\ 
\label{eq:esav}
&=&\frac{1}{\Omega^N}\int\cdots\int_\Omega\Psi(\xx;\rr_1,\rr_2,\ldots,\rr_N)\,
d\rr_1d\rr_2\ldots d\rr_N.
\eq
Clearly if we can neglect the interaction between the $N$ bosons and the
particle we will have $\widetilde{\Psi}\propto\psi$, but this is not
true anymore in presence of a coupling between the particle and the
scattering centers.

The Hamiltonian of the whole system
\bq \label{eq:Hf}
H=\frac{p^2}{2m}+\sum_j\hbar\omega_j+\sum_j v(\xx-\rr_j),
\eq
where $\pp$ is the momentum of the particle, may also be rewritten, 
at $T<T_C$, as the following operator
\footnote{Here we use the hat only for the operators acting on the 
scattering centers vacuum and not on the operators of the particle.}
\bq \nonumber
\hat{H}&=&\frac{p^2}{2m}+\sum_j\hbar\omega_j 
\hat{b}_j^\dagger \hat{b}_j+
\int_\Omega \sum_j v(\xx-\rr')\hat{b}_j^\dagger \hat{b}_j\,d\rr'\\ \label{eq:Ho}
&=&\frac{p^2}{2m}+
\int_\Omega \sum_j v(\xx-\rr')\hat{b}_j^\dagger \hat{b}_j\,d\rr',
\eq
where $\hat{b}_j^\dagger$ is the creation operator of scattering center 
$j$ such that the number operator $\hat{n}_j=\hat{b}_j^\dagger \hat{b}_j$, 
for example, in his position $\rr$ representation acts as follows 
\red{$\langle \rr|\hat{n}_j|0\rangle=\delta^{(d)}(\rr-\rr_j)$}, with $|0\rangle$ the 
vacuum defined as the state that is annihilated by the destruction 
operator $\hat{b}_j$. In the second equality of Eq. (\ref{eq:Ho}) we
explicitly used the fact that below the Bose-Einstein critical
temperature the $N$ boson scattering centers are all in their condensed 
phase at $\omega_j=0$ for all $j$. Here we are thinking at the 
condensed $N$ scattering centers as being independent one from 
the other and non interacting among themselves so that Eq. 
(\ref{eq:phibec}) may be rewritten as
\red{
\bq
\Phi(\{\rr_j\})=\langle\rr_1,\rr_2,\ldots,\rr_N|\left(\prod_j\int_\Omega\hat{n}_j
\,d\rr_j'\right)|0\rangle.
\eq
}
Then the Hamiltonian of Eq. (\ref{eq:Hf}) is
the result of the action of the operator $\hat{H}$ of Eq.
(\ref{eq:Ho}) on the vacuum. Written in the form of Eq. (\ref{eq:Ho})
the Edwards Hamiltonian resembles the polaron Hamiltonian 
\cite{Fantoni12d,Fantoni13a,Fantoni25z}. 
In particular the recipe of Eq. (\ref{eq:esav})
of washing out the degrees of freedom of the scattering centers by 
averaging on their positions finds its justification in the need of
a polaron description \cite{Fantoni25z}.

In the next Section \ref{sec:lax-canonical} we will choose a particular 
form for $v$ and will see how the averaging recipe of Eq. (\ref{eq:esav})
favors localization.

\section{The Lax model with canonical quantization}
\label{sec:lax-canonical}

Lax and Phillips choose $d=1$ and $v(x)=g\delta(x)$ \cite{Lax1958} with $g$
the coupling constant between the particle and the boson scattering centers
in their condensed phase so that the kinetic energy of the bosons can
be neglected in the Hamiltonian of Eq. (\ref{eq:Hf}) as in 
Eq. (\ref{eq:Ho})
\bq \label{eq:Hlax}
H=-\frac{1}{2m}\frac{\partial^2}{\partial x^2}+g\sum_{j=1}^N\delta(x-r_j),
\eq
where we set $\hbar=1$ and we may order $0<r_1<r_2<\ldots<r_N<L$ the 
positions of the $N$ scattering centers in the periodic segment 
$[0,L[$ (a circle). \green{Therefore the scattering centers have a 
linear number density $\rho=N/L$}. We will then define $r_0=r_N$. In 
Appendix A of Ref.
\cite{Rose2022} it was shown that for an arbitrary choice of the
$N$ positions $\{r_i\}$, at large $g$, the eigenstates of the 
Hamiltonian (\ref{eq:Hlax}), $\Psi_n(x;r_1,r_2,\ldots,r_N)$, become 
localized on the circle. Here 
$H\Psi_n(x;\{r_j\})=E_n(\{r_j\})\Psi_n(x;\{r_j\})$ where their 
eigenvalues, $E_n(\{r_j\})=k_n^2(\{r_j\})/2m$, are determined from 
the periodic boundary condition $\Psi_n(0;\{r_i\})=\Psi_n(L;\{r_i\})$, 
where $n=0$ corresponds to the ground state and integers $n>0$ to the 
excited states (see our Appendix \ref{app:A}).

We here want to see whether the localization is {\sl robust} against 
switching on of disorder, i.e. averaging over stochastic choices of 
the $\{r_i\}$.
In a Monte Carlo spirit \cite{Kalos-Whitlock} we will then generate
$MN$ pseudo random numbers $\{r_i\}^k$, with $k=1,2,\ldots,M$, ordered 
within $[0,L[$ and according to Eq. (\ref{eq:esav}) we will measure
\bq \nonumber
\widetilde{\Psi}_n(x)
&=&\langle\Psi_n(x;r_1,r_2,\ldots,r_N)\rangle\\ \label{eq:esmc}
&=&\frac{1}{M}\sum_{k=1}^M\Psi_n(x;r_1^k,r_2^k,\ldots,r_N^k),
\eq
at fixed $n$. Together with 
\bq \nonumber
\widetilde{E}_n&=&\langle E_n(r_1,r_2,\ldots,r_N)\rangle\\ \label{eq:evmc}
&=&\frac{1}{M}\sum_{k=1}^ME_n(r_1^k,r_2^k,\ldots,r_N^k).
\eq

\begin{figure}[htbp]
\begin{center}
\includegraphics[width=8cm]{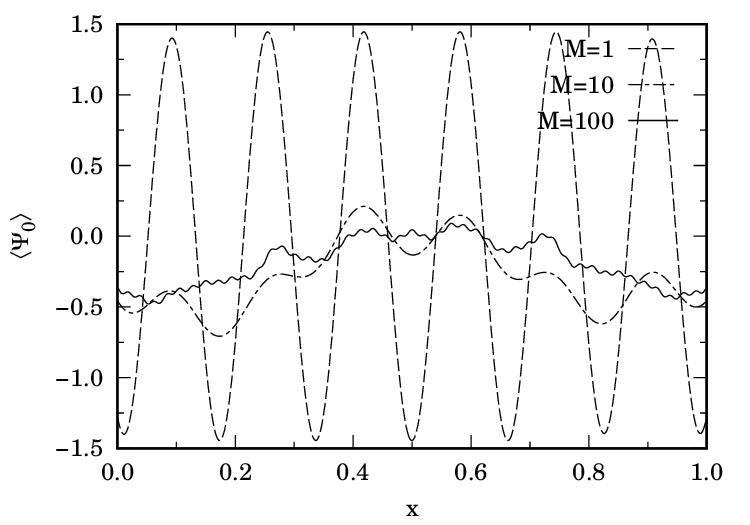}
\includegraphics[width=8cm]{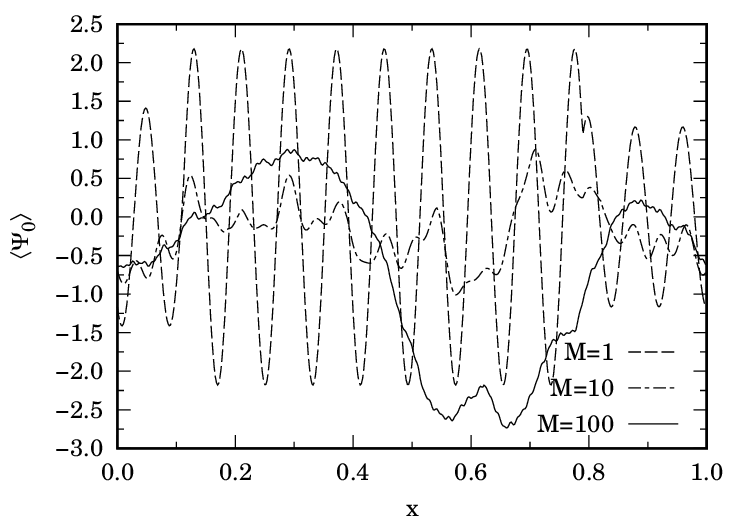}
\end{center}  
\caption{We show the ground state from Eq. (\ref{eq:esmc}) for three 
choices of increasing $M$ and $N=3, L=1$. Low $g=1$ in the left panel and 
high $g=100$ in the right panel. From Eq. (\ref{eq:evmc}) the 
resulting ground state energy is as follows: for $g=1\longrightarrow$
($\widetilde{E}_0=741.956$ for $M=1$, $\widetilde{E}_0=148.428$ for $M=10$, 
and $\widetilde{E}_0=613.504$ for $M=100$) and for $g=100\longrightarrow$
($\widetilde{E}_0=3027.62$ for $M=1$, $\widetilde{E}_0=2343.16$ for $M=10$, 
and $\widetilde{E}_0=3698.51$ for $M=100$).} 
\label{fig:esmc}
\end{figure}

In Fig. \ref{fig:esmc} we show the ground state for three choices of 
increasing $M$ at low $g=1$ and high $g=100$ for $N=3$. As we can see 
the randomness in the positions of the scattering centers produces 
localization. And moreover increasing the coupling $g$ with the scattering 
centers increases the localization of the averaged ground state
wavefunction. From the figure we also see how at small $g$ the amplitude 
of the localized averaged ground state wavefunction is much smaller than 
the amplitude of the unaveraged ground state, but the opposite behavior 
is observed at high coupling $g$. 

\begin{figure}[htbp]
\begin{center}
\includegraphics[width=8cm]{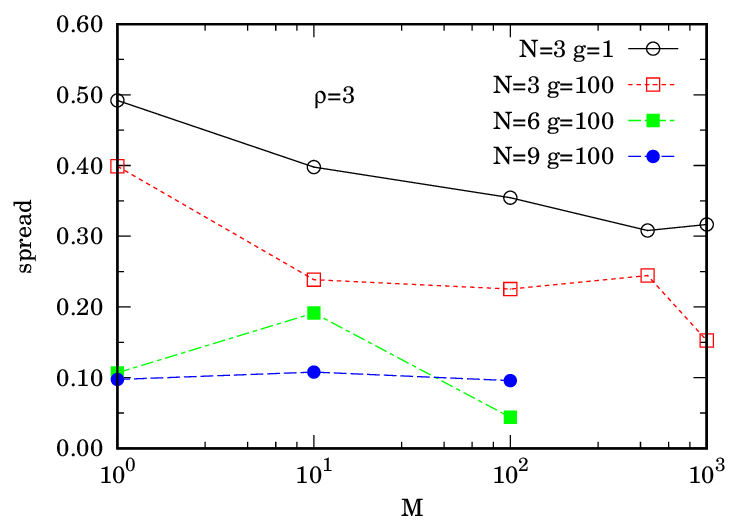}
\end{center}  
\caption{\green{We show the ground state wavefunction spread 
of Eq. (\ref{eq:spread}) as a function of the Monte Carlo steps 
$M$ for $m=1,\rho=3$ and the four cases $N=3, L=1, g=1$, $N=3, L=1, g=100$,
$N=6, L=2, g=100$, and $N=9, L=3, g=100$.}} 
\label{fig:loc}
\end{figure}

\green{
As a measure of the localization property we compute the spread 
in the particle averaged ground state as
\bq \label{eq:spread}
\mbox{spread}=\frac{\int_0^L|\tilde{\Psi}_0|^2\,dx}{G_{\rm max}L}\in [0,1],
\eq
where $G_{\rm max}$ is the global maximum of $|\tilde{\Psi}_0|^2$ in 
$[0,L[$. A completely delocalized ground state has a spread equal to 1.
In Fig. \ref{fig:loc} we plot the spread as a function of $M$ for 
$m=1,\rho=3$ and the four cases $N=3, L=1, g=1$, $N=3, L=1, g=100$,
$N=6, L=2, g=100$, and $N=9, L=3, g=100$. From the figure we see how
as $M$ increases the localization increases. Moreover the strongly 
coupled particle is more localized than the weakly coupled one. This 
localization property remains valid approaching the thermodynamic limit
where $N\to\infty$ and $L\to\infty$ at constant $\rho$. In particular we 
see from the figure that increasing $N$ at constant density increases
the localization.
}   

\subsubsection{Dynamics}

The dynamic evolution of the $n$-th eigenstate is as usual
\bq
\Psi_n(x,t;\{r_j\})=e^{-iE_n(\{r_j\})t}\Psi_n(x;\{r_j\}),
\eq
according to the time, $t$, dependent Schr\"odinger equation
$i\partial\Psi/\partial t=H\Psi$. 

We may then initially think at the following severe short times 
Monte Carlo approximation
\bq \nonumber
\widetilde{\Psi}_n(x,t)&=&\langle\Psi_n(x,t;\{r_j\})\rangle\\ \nonumber
&=&\langle e^{-iE_n(\{r_j\})t}\Psi_n(x;\{r_j\})\rangle\\ \nonumber
&\approx&\langle e^{-iE_n(\{r_j\})t}\rangle\langle\Psi_n(x;\{r_j\})\rangle\\
\nonumber
&\approx& e^{-i\langle E_n(\{r_j\})\rangle t}
\langle\Psi_n(x;\{r_j\})\rangle\\ \label{eq:dyn-severe}
&=& e^{-i\widetilde{E}_nt}\widetilde{\Psi}_n(x),
\eq
where the second approximation may be justified for a small
times evolution. On the other hand, the first approximation is a 
rather severe one. Clearly it holds exactly only for $N=0$.
But from the approximation (\ref{eq:dyn-severe}) follows that
$|\widetilde{\Psi}_n(x,t)|^2=|\widetilde{\Psi}_n(x)|^2$
independent of time.

We then see that in order to have a time dependent probability
distribution it is essential to stick to the definition
\footnote{Here we could have defined
$\calp\propto\langle|\Psi_n(x,t;\{r_j\})|^2\rangle$ 
or even
$\calp\propto\langle|\Psi_n(x,t;\{r_j\})|\rangle^2$, as well. 
All three definitions have equal dignity. Our choice 
(\ref{eq:dyn1}) give importance to the naked wavefunction.}
\bq \label{eq:dyn1}
\calp(x,t)&=&\frac{|\widetilde{\Psi}_n(x,t)|^2}
{\int_0^L|\widetilde{\Psi}_n(x,t)|^2\,dx}\\ \label{eq:dyn2}
\widetilde{\Psi}_n(x,t)&=&\langle\Psi_n(x,t;\{r_j\})\rangle.
\eq

\begin{figure}[htbp]
\begin{center}
\includegraphics[width=8cm]{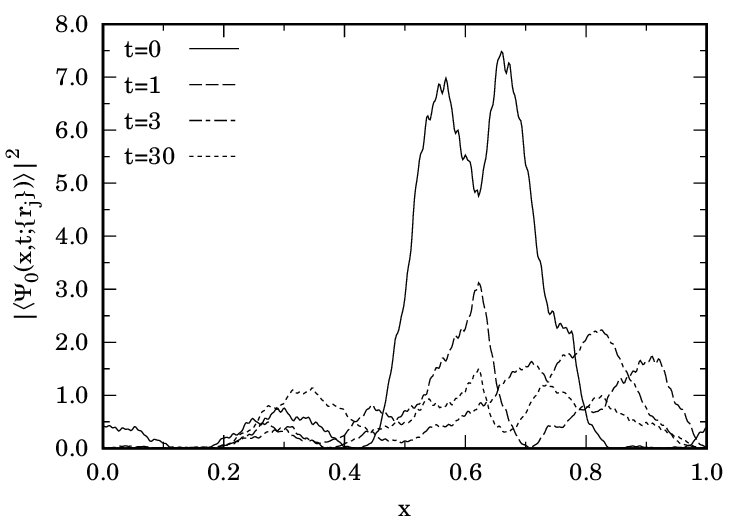}
\end{center}  
\caption{We show the time evolution of $|\widetilde{\Psi}_0(x,t)|^2$
from Eq. (\ref{eq:dyn2}) for $N=3, L=1, g=100, M=100$ at $t=0,1,3,30$.} 
\label{fig:esmct}
\end{figure}

In Fig. \ref{fig:esmct} we show the time evolution of 
$|\widetilde{\Psi}_0(x,t)|^2$ for the ground state $n=0$ for 
$N=3, L=1, g=100$ with $M=100$. As we can see from the figure 
the probability density initially localized around $x\sim 0.6$ 
faints out at large times. 

We may be interested in solving Schr\"odinger equation with an
arbitrary initial condition
\bq
\Psi(x,0;\{r_j\})=\green{I}(x;\{r_j\}),
\eq
This is a much harder problem requiring a path integral for the
propagator \cite{Feynman-Hibbs}
\bq
T(x,t|x_0,0;\{r_j\})&=&\int_{[0,t]\times[0,L[}^{\rule[.1cm]{1.5cm}{.02cm}}
\!\!\!\!\int_{x(0)=x_0}^{x(t)=x} 
e^{iS[x,\dot{x},\tau;\{r_j\}]}\,\cald x(\tau),\\
T(x,0|x_0,0)&=&\delta(x-x_0),
\eq
where we denoted with a dot a time derivative and the action functional
\bq
S[x,\dot{x},\tau;\{r_j\}]=\int_0^\tau\left(\frac{1}{2}m\dot{x}^2-
g\sum_j\delta(x-r_j)\right)\,dt,
\eq
where the integrand is the Lagrangian function. Then
\bq
\Psi(x,t;\{r_j\})=\int_0^L 
T(x,t|x',0;\{r_j\})\green{I}(x';\{r_j\})\,dx'.
\eq
For example, if the particle is initially localized at $x_0\in[0,L[$,
with $\green{I}(x;\{r_j\})=\delta(x-x_0)$, then its time evolution will be given 
by $\Psi(x,t;\{r_j\})=T(x,t|x_0,0;\{r_j\})$. In the small time limit
we will have $\Psi(x,t;\{r_j\})\approx e^{iS[x,\dot{x},t;\{r_j\}]}$
\cite{LandauQM}.

Thinking at the particle as an electron the instantaneous electrical 
current in the circle with impurities would then be 
\bq
\cali(t)=e\frac{d}{dt}\left(
\frac{\int_0^L|\langle\Psi(x,t;\{r_j\})\rangle|^2x\,dx}
{\int_0^L|\langle\Psi(x,t;\{r_j\})\rangle|^2\,dx}\right),
\eq
where $e$ is the charge of the electron.

\section{The Lax model with affine quantization}
\label{sec:lax-affine}

It would then be very interesting to repeat the calculation within
{\sl affine quantization} (see Appendix A in Ref. \cite{Fantoni23b}) 
where
\bq \label{eq:Haff}
\calh &=&\left[-\frac{1}{2m}\frac{\partial^2}{\partial x^2}+
v^{\rm aff}(x)\right]+g\sum_{j=1}^N\delta(x-r_j),\\
v^{\rm aff}(x)&=&\frac{3\left[\sum_j(x-a_j)\right]^2+N\sum_j[b_j^2-(x-a_j)^2]}
{2m\left\{\sum_j[b_j^2-(x-a_j)^2]\right\}^2},~~~
b_j=\frac{r_j-r_{j-1}}{2},~~~
a_j=\frac{r_j+r_{j-1}}{2},
\eq
where the affine effective potential terms $v_j^{\rm aff}$ are 
the results of adopting the affine {\sl dilation} operator $\cald$, 
in place of the canonical momentum operator $p$,
\bq
\cald=\sum_j\left\{p^\dagger[b_j^2-(x-a_j)^2]+
[b_j^2-(x-a_j)^2]p\right\}/2,
\eq
in the affinely quantized Hamiltonian
\bq
\calh=\cald\left\{\sum_j[b_j^2-(x-a_j)^2]\right\}^{-2}\cald + 
g\sum_{j=1}^N\delta(x-r_j),
\eq
where the effective affine potential insures that the 
particle will live in any of the segments $(x-a_j)^2<b_j^2$, i.e. 
$x\in ]r_{j-1},r_j[$, tunneling from one to the other 
through the delta function. From Eq. (\ref{eq:Haff}) follows the
eigenstate Schr\"odinger equation for an eigenvalue $E$
\bq
\left\{\sum_j[b_j^2-(x-a_j)^2]\right\}^2\left(
\frac{\partial^2\Psi}{\partial x^2}+2mE\Psi\right)
=\left\{3\left[\sum_j(x-a_j)\right]^2+N\sum_j[b_j^2-(x-a_j)^2]\right\}\Psi,
\eq
where the delta function terms vanished. This decouples the particle
from the bosonic scattering centers. We conclude that in affine
quantization the eigenstates must have a continuous first derivative, 
i.e. they are smooth functions, and that the localization
holds independently from $g$! Again we expect that making the 
scattering centers stochastic will increase the localization of the
particle ground state wave function averaged over the disorder.
But this is a complicated computational problem that we will leave 
open for the future.

\section{Conclusions}
\label{sec:concl}

In this work we find numerically the ground state energy and wavefunction
of a particle, in a one dimensional periodic square well, coupled 
through a pair Dirac delta function potential with $N$ scattering centers 
situated at fixed random positions within the well as in the Edwards-Lax 
model. We show that the wavefunction becomes localized for $N>1$ and 
for any value of the coupling constant $g$. 

Averaging the ground state over the positions of the scattering
centers chosen as pseudo random numbers uniformly distributed within the
well we show that the localization increases.

\green{We estimated the degree of localization showing that the particle
averaged ground state becomes more localized as the number of samples 
of random scattering centers increases, as the particle is strongly 
coupled with the scattering centers, and as the number of scattering 
centers increases at constant number density.}

We also show that the time dependent averaged ground state wavefunction 
becomes more localized at intermediate times and then faints out at large
times.

We conclude studying the same system in affine quantization and show 
that the resulting problem is independent of the coupling constant $g$.

Even if we are still far from a unified theory of localization in 
quantum stochastic mechanics, we hope that the present study could 
represent a useful addition to the physics literature on the 
localization phenomenon.

\appendix
\section{Determination of the eigenstates and the eigenvalues of Lax model}
\label{app:A}

Choose $N$ ordered random real numbers $0<r_1<r_2<\ldots<r_N<L$. 
Define $r_0=0$. Solve the Schr\"odinger equation for 
$\psi(x)=\Psi(x;\{r_j\})$
\bq \label{eq:SchLax}
E\psi(x)=-\frac{1}{2m}\psi''(x)+g\sum_{j=1}^N\delta(x-r_j)\psi(x)
\eq
with the {\sl transfer matrix method}. As usual we denote with a prime a
derivative respect to $x$. For $g>0$ multiplying this equation
by $\psi^\star$ and integrating over $x\in [0,L[$ follows immediately 
$E>0$. On any subinterval $]r_j,r_{j+1}[$ the wavefunction takes the form
\bq \label{eq:psi}
\psi(x)=A_j\sin(kx+\varphi_j),
\eq
with $A_i$ and $\varphi_i$ an amplitude and a phase and $E=k^2/2m$. 
Integrating Eq. (\ref{eq:SchLax}) around $r_j$ and using the continuity of
$\psi(x)$ follows
\bq \label{eq:psip}
\psi'(r_j^+)-\psi'(r_j^-)=2mg\psi(r_j).
\eq
Constructing the column vector $C(x)=(\psi'(x),k\psi(x))^T$, the conditions of 
Eqs. (\ref{eq:psi}) and (\ref{eq:psip}) can be rewritten as
\bq
C(r_{j+1}^-)=R[k(r_{j+1}-r_j)]C(\green{r}_j^+),~~~C(\green{r}_j^+)=TC(\green{r}_j^-),
\eq
respectively, where
\bq
R(\theta)=\left(\begin{array}{cc}
\cos\theta & -\sin\theta\\
\sin\theta & \cos\theta
\end{array}\right),~~~
T=\left(\begin{array}{cc}
1 & 2mg/k\\
0 & 1
\end{array}\right).
\eq
Hence $C(L)=\green{\calm}C(0)$, with
\bq
\green{\calm}=R[k(L-r_N)]TR[k(r_N-r_{N-1})]T\cdots TR[k(r_1-r_0)]T.
\eq
The spectrum is fixed by the boundary condition $\psi(0)=\psi(L)$. 
For example for periodic boundary conditions $\green{\calm}=1$ and 
${\rm tr}\green{\calm}=2$,
which can be solved with the Newton-Raphson (NR) method. We found the
ground state energy by choosing as seed for the NR algorithm $k=0^+$.

Once the spectrum is determined one relates $A_{j+1}$ and $\varphi_{j+1}$
to $A_j$ and $\varphi_j$ using the continuity of $\psi(x)$ and the jump
condition (\ref{eq:psip}). Thus, by recursion, any $A_j$ and $\varphi_j$ 
for $j=0,1,\ldots,N$ can
be expressed in terms of $A_0$ and $\varphi_0$. We determine $A_0$ in terms of
$\varphi_0$ requiring that $\psi(x)$ is normalized over the whole well $[0,L[$.
Finally, we determine $\varphi_0$ numerically from the boundary condition
$\psi(0)=\psi(L)$ with the NR method with any seed.

\section*{Author declarations}

\subsection*{Conflicts of interest}
None declared.

\subsection*{Data availability}
The data that support the findings of this study are available from the 
corresponding author upon reasonable request.

\subsection*{Funding}
None declared.

\bibliography{el}

@Article{Lax1958,
author = {M. Lax and J. C. Phillips},
title = {{One-dimensional impurity bands}},
journal = {Phys. Rev.},
year = {1958},
OPTkey = {•},
volume = {110},
OPTnumber = {•},
pages = {41},
OPTmonth = {•},
OPTnote = {•},
OPTannote = {•},
doi={10.1103/PhysRev.110.41}
}

@Article{Edwards1958,
author = {S. F. Edwards},
title = {{A new method for the evaluation of electric conductivity in metals}},
journal = {Philosophical Magazine},
year = {1958},
OPTkey = {•},
volume = {3},
OPTnumber = {•},
pages = {33},
OPTmonth = {•},
OPTnote = {•},
OPTannote = {•},
doi={10.1080/14786435808243244}
}

@Article{Rose2022,
author = {F. Rose and R. Schmidt},
title = {{Disorder in order: Localization without randomness in a cold-atom system}},
journal = {Phys. Rev. A},
year = {2022},
OPTkey = {•},
volume = {105},
OPTnumber = {•},
pages = {013324},
OPTmonth = {•},
OPTnote = {•},
OPTannote = {•},
doi={10.1103/PhysRevA.105.013324}
}

@Article{Anderson1958,
author = {P. W. Anderson},
title = {{Absence of Diffusion in Certain Random Lattices}},
journal = {Phys. Rev.},
year = {1958},
OPTkey = {•},
volume = {109},
OPTnumber = {•},
pages = {1492},
OPTmonth = {•},
OPTnote = {•},
OPTannote = {•},
doi={10.1103/PhysRev.109.1492}
}

@Article{Frohlich1950,
author = {H. Fr\"ohlich and H. Pelzer and S. Zienau},
title = {{Properties of slow electrons in polar materials}},
journal = {Phil. Mag.},
year = {1950},
OPTkey = {•},
volume = {41},
OPTnumber = {•},
pages = {221},
OPTmonth = {•},
OPTnote = {•},
OPTannote = {•},
doi={10.1080/14786445008521794}
}

@Book{LandauQM,
  author = 	 {L. D. Landau and E. M. Lifshitz},
  ALTeditor = 	 {},
  title = 	 {Quantum Mechanics. Non-relativistic Theory},
  publisher = 	 {Pergamon Press Ltd},
  year = 	 {1977},
  OPTkey = 	 {},
  volume = 	 {3},
  OPTnumber = 	 {},
  OPTseries = 	 {},
  address = 	 {Oxford},
  OPTedition = 	 {},
  OPTmonth = 	 {},
  note = 	 {Course of Theoretical Physics. Section \S 6},
  OPTannote = 	 {}
}

@book{Feynman-Hibbs,
   title =     {{Quantum Mechanics and Path Integrals}},
   author =    {Richard P. Feynman and Albert R. Hibbs and Daniel F. Styer},
   publisher = {Dover Publications},
   OPTisbn =      {9780486477220,0486477223,2010004550},
   year =      {2010},
   series =    {},
   edition =   {Emended},
   volume =    {},
   OPTurl =       {http://gen.lib.rus.ec/book/index.php?md5=132E2E46DF4093FED8D9F84AEFFABBD7},
   note = {page 292-293}
}

@Book{Kalos-Whitlock,
  author = 	 {M. H. Kalos and P. A. Whitlock},
  ALTeditor = 	 {},
  title = 	 {Monte Carlo Methods},
  publisher = 	 {John Wiley \& Sons Inc.},
  year = 	 {1986},
  OPTkey = 	 {},
  OPTvolume = 	 {},
  OPTnumber = 	 {},
  OPTseries = 	 {},
  address = 	 {New York},
  OPTedition = 	 {},
  OPTmonth = 	 {},
  OPTnote = 	 {},
  OPTannote = 	 {}
}

@Article{Fantoni25z,
  author  = {Riccardo Fantoni},
  journal = {under submition},
  title   = {{Polaron versus Anderson Localization}},
  year    = {2025},
  note={ArXiV:2510.16434}
}

@Article{Fantoni23b,
  author  = {J. R. Klauder and R. Fantoni},
  journal = {Axioms},
  title   = {{The Magnificent Realm of Affine Quantization: valid results for particles, fields, and gravity}},
  year    = {2023},
  pages   = {911},
  volume  = {12},
  doi     = {10.3390/axioms12100911},
}

@Article{Fantoni13a,
  author  = {R. Fantoni},
  journal = {Physica B},
  title   = {{Low temperature acoustic polaron localization}},
  year    = {2013},
  pages   = {112},
  volume  = {412},
  doi     = {10.1016/j.physb.2012.12.032},
}

@Article{Fantoni12d,
  author  = {R. Fantoni},
  journal = {Phys. Rev. B},
  title   = {{Localization of acoustic polarons at low temperatures: A path integral Monte Carlo approach}},
  year    = {2012},
  pages   = {144304},
  volume  = {86},
  doi     = {10.1103/PhysRevB.86.144304},
}

\end{document}